\documentclass[fleqn,twoside]{article}
\usepackage{espcrc2,psfig}

\newlength{\figwidth}
\renewcommand{\vec}[1]{\mbox{\boldmath $#1$}}

\title{Charmed hadron physics in quenched anisotropic lattice QCD%
       \thanks{Poster presented by H. Matsufuru.}}

\author{
J. Harada\address{%
    Department of Physics, Hiroshima University,
    Higashi-hiroshima 739-8526, Japan \vspace{-0.25cm}},
H. Matsufuru\address{%
    Yukawa Institute for Theoretical Physics, Kyoto University,
    Kyoto 606-8502, Japan \vspace{-0.25cm} },
T.~Onogi$^{\rm b}$
and
A. Sugita\address{%
    Research Center for Nuclear Physics, Osaka University,
    Ibaraki 567-0047, Japan } }

\begin{document}

\begin{abstract}
We investigate the anisotropic lattice with $O(a)$ improved
quark action as a candidate of framework in which we can treat
both the heavy and light quark region in the same manner and
systematically reduce the systematic uncertainties.
To examine applicability of anisotropic lattice, we calculate
the charmed meson spectrum and decay constants in quenched
approximation. We find consistent result with 
most advanced results on isotropic lattices.
\end{abstract}

\maketitle

\section{Introduction}
  \label{sec:introduction}

Recent experimental progress in flavor physics to look for 
the effect of new physics strongly requires precise theoretical 
predictions from the standard model. However, model independent 
calculation of hadronic matrix elements is difficult because of 
nonperturbative nature of QCD. The lattice QCD simulation is one 
of most promising approaches in which the systematic uncertainties 
can be reduced systematically
\cite{review}.
The ultimate goal of this work is to construct a framework
for lattice calculations of hadronic matrix elements in a few percent
systematic accuracy, as required by the experiments in progress.

In a lattice calculation of heavy quarks such as the charm and bottom,
one need to either avoid or control the large discretization error
of $O((m_Q a)^n)$.  Although the effective theoretical approaches
\cite{NRQCD,EKM97} successfully describe the matrix elements 
with heavy quarks  within 10\% accuracy, further improvement
of accuracy is difficult due to perturbative error.
The nonperturbative renormalization method \cite{NRimp} are not
directly applicable to a determination of the coefficient of $O(a)$ 
improving term in the region of $m_Q\simeq a^{-1}$ because of 
a strong mass dependence. Therefore a new framework is desired for which a
systematic nonperturbative improvement can be performed.

The anisotropic lattice, on which the temporal lattice spacing
$a_{\tau}$ is finer than the spatial one $a_\sigma$,
is a candidate of such a framework.
A high temporal lattice cutoff may enable a relativistic
treatment of heavy quark while keeping computational requirement
tractable.
The $O(a)$ improvement coefficients determined in the light quark
region in good precision may directly be available for a heavy quark
with $m_Q\ll a_{\tau}^{-1}$.
Whether these observations practically hold should be
examined numerically, as well as in the perturbation theory.

In this study, we examine the applicability of anisotropic
lattice approach numerically by computing the charmed meson masses 
and decay constants.
Although presently achieved accuracy is far from our goal,
the result is encouraging for further development in this direction.
This paper shows our preliminary result without technical details
and evaluation of the systematic uncertainties,
which are fully discussed in other publications
\cite{Aniso01a,Aniso01b,Aniso02a}.

\section{Anisotropic lattice quark action}
\label{sec:formulation}

The quark action we adopt has the following structure
in the hopping parameter form \cite{Aniso01a,Ume01}:
{\small
\begin{eqnarray}
 S_F  \!\!\!&=&\!\!\! \sum_{x,y} \bar{\psi}(x) \left\{
 \delta_{x,y}
   - \kappa_{\tau} \left[ \ \ (1-\gamma_4)U_4(x)\delta_{x+\hat{4},y} \right.
 \right.
 \nonumber \\
 & &  \hspace{1cm}
      + \left. (1+\gamma_4)U_4^{\dag}(x-\hat{4})\delta_{x-\hat{4},y} \right]
 \nonumber \\
 & & \hspace{-0.8cm}
    -  \kappa_{\sigma} {\textstyle \sum_{i}}
         \left[ (r\!-\!\gamma_i) U_i(x) \delta_{x+\hat{i},y}
    (r\!+\!\gamma_i)U_i^{\dag}(x\!-\!\hat{i})\delta_{x-\hat{i},y} \right]
 \nonumber \\
 & & \hspace{-0.0cm}
    -  \kappa_{\sigma} c_E
             {\textstyle \sum_{i}} \sigma_{4i}F_{4i}(x)\delta_{x,y}
 \nonumber \\
 & & \hspace{-0.0cm}
 \left.
    - r \kappa_{\sigma} c_B
             {\textstyle \sum_{i>j}} \sigma_{ij}F_{ij}(x)\delta_{x,y},
   \right\} \psi(y), 
 \label{eq:action}
\end{eqnarray}
}
where $\kappa_{\sigma}$ and  $\kappa_{\tau}$ 
are the spatial and temporal hopping parameters, and are related
to the bare quark mass and bare anisotropy as given below.
$r$ is the spatial Wilson parameter, and $c_E$ and $c_B$ are the
coefficients of clover terms which remove $O(a)$ errors.
Although the explicit Lorentz symmetry is removed due to the
anisotropy, in principle it can be restored for physical observables
up to errors of $O(a^2)$ by proper tuning of
 $\kappa_{\sigma}/\kappa_{\tau}$, $r$, $c_E$ and $c_B$
for a given $\kappa_{\sigma}$.
The action is constructed in accord with the Fermilab approach
\cite{EKM97} and hence applicable to an arbitrary quark mass.
However, a mass dependent tuning of parameters is difficult
beyond perturbation theory.
This may be circumvented by taking $a_\tau^{-1}\gg m_Q$, with which
the mass dependence of parameters are expected to be small,
and it suffices with those values in the chiral limit.

In present study, we vary only two parameters $\kappa_\sigma$
and $\kappa_\tau$ with fixed other parameters.
We set the Wilson parameter as $r=1/\xi$ and
the clover coefficients as the tadpole-improved tree-level
values, $c_E= 1/u_{\sigma} u_{\tau}^2$, and $c_B = 1/u_{\sigma}^3$.
The tadpole improvement \cite{LM93} is achieved
by rescaling the link variable as
$U_i(x) \rightarrow U_i(x)/u_{\sigma}$ and  $U_4(x) \rightarrow
U_4(x)/u_{\tau}$, with the mean-field values of the spatial 
and temporal link variables, $u_{\sigma}$ and $u_{\tau}$,
respectively.
Instead of $\kappa_\sigma$ and $\kappa_\tau$,
we introduce $\kappa$ and $\gamma_F$ as
\begin{eqnarray}
\frac{1}{\kappa} &\equiv& \frac{1}{\kappa_{\sigma} u_\sigma}
     - 2(\gamma_F+3r-4)
    = 2(m_0 \gamma_F  +4) , \nonumber \\
\gamma_F  &\equiv&
              \kappa_\tau u_\tau / \kappa_\sigma u_\sigma .
 \label{eq:kappa}
\end{eqnarray}
The former plays the same role as on the isotropic lattice, and
the latter corresponds to the bare anisotropy.

On an anisotropic lattice, one must tune the parameters so that the
anisotropy of quark field, $\xi_F$, equals that of the gauge
field, $\xi_G$:
\begin{equation}
 \xi_F(\beta,\gamma_G;\kappa,\gamma_F)
 = \xi_G(\beta,\gamma_G;\kappa,\gamma_F)
= \xi .
\label{eq:xi}
\end{equation} 
Although $\xi_G$ and $\xi_F$ are in general functions of both of
gauge parameters ($\beta$, $\gamma_G$) and quark parameters
($\kappa$, $\gamma_F$),
on a quenched lattice one can determine $\xi_G=\xi$ independently of
$\kappa$ and $\gamma_F$, and then tune $\gamma_F$ so that a certain
observable satisfies the condition (\ref{eq:xi}).
In this work, we define $\xi_F$ through the relativistic dispersion
relation of meson,
\begin{equation}
E^2 (\vec{p}) = m^2 + \vec{p}^2 / \xi_F^2 + O[(\vec{p}^2)^2],
\label{eq:DR1}
\end{equation}
for calibration.
In the above expression, the energy and mass $E$ and $m$ 
are in temporal lattice units while the momentum $\vec{p}$ is
in spatial lattice units.
$\xi_F$ converts the momentum into that in temporal lattice units.

\section{Numerical simulation}
\label{sec:numerical}

Numerical simulations are performed on two quenched lattices
of sizes $16^3\times 128$ and $20^3\times 160$
with the Wilson plaquette action at $\beta=5.95$ and $6.10$,
respectively, with the renormalized anisotropy $\xi=4$
\cite{Aniso01b}.
The values of bare anisotropy are chosen according to
a numerical result performed in one percent accuracy
in \cite{Kla98}.
The lattice scales are set by $K^*$ meson mass, and result in
 $a_\sigma^{-1}=1.525(27)$ GeV and $1.817(22)$ GeV for 
 $\beta=5.95$ and $6.10$, respectively.
These values deviate to about 10\% from scales determined
with other physical quantity, say hadronic radius $r_0$,
and represent the effect in neglecting dynamical quarks.
The mean field values $u_\tau$ and $u_\sigma$
are obtained in the Landau gauge \cite{Ume01}.

The calibration was done in the quark mass region below around charm
quark mass in \cite{Aniso01b}.
It was found that the quark mass dependence of bare anisotropy
$\gamma_F^*$, with which $\xi_F=\xi$ holds, is actually small,
and $\gamma_F^*$ is well fitted to a linear form in $m_q^2$,
\begin{equation}
 \frac{1}{\gamma_F^*}(m_q) = \zeta_0 + \zeta_2 m_q^2,
 \hspace{0.3cm}
 m_q = \frac{1}{2 \xi} \left( \frac{1}{\kappa} 
    - \frac{1}{\kappa_c} \right) .
\label{eq:fit_calib}
\end{equation}
The fit results in the values ($\zeta_0, \zeta_2, \kappa_c$) = 
(0.2490(8), 0.189(15), 0.12592(8)) at $\beta=5.95$, and
(0.2479(9), 0.143(14), 0.12558(9)) at $\beta=6.10$.
These values are determined in 1\% statistical accuracy,
while in the chiral limit additional 1\% error exists
due to the form of extrapolation.
It was also shown that the systematic discretization errors
decrease toward the continuum limit.
The light hadron spectra with the obtained parameters are
consistent with previous works on isotropic lattices.

\begin{table}
\caption{
The heavy-light meson masses and decay constants for physical
quark masses in physical units.}
\begin{center}
\begin{tabular}{ccccc}
\hline\hline
 {[GeV]} & $\beta=5.95$ & $\beta=6.10$ \\
\hline
 $m_{D_s}$           & 1.9740(28)& 1.9760(27) \\
 $m_{D^*}-m_{D}$     & 0.1018(85)& 0.1022(74) \\
 $m_{D^*_s}-m_{D_s}$ & 0.0980(42)& 0.0921(39) \\
\hline
 $f_\pi$             & 0.1655(35)& 0.1462(35) \\
 $f_K$               & 0.1868(29)& 0.1665(30) \\
 $f_D$               & 0.2509(64)& 0.2245(58) \\
 $f_{D_s}$           & 0.2863(40)& 0.2570(37) \\
\hline
 $f_D/f_\pi$         & 1.516(44) & 1.536(49) \\
 $f_{D_s}/f_D$       & 1.141(15) & 1.145(16) \\
\hline\hline
\end{tabular}
\end{center}
\vspace{-1.0cm}
\label{tab:decayconst_latt}
\end{table}

\begin{figure}[tb]
\psfig{file=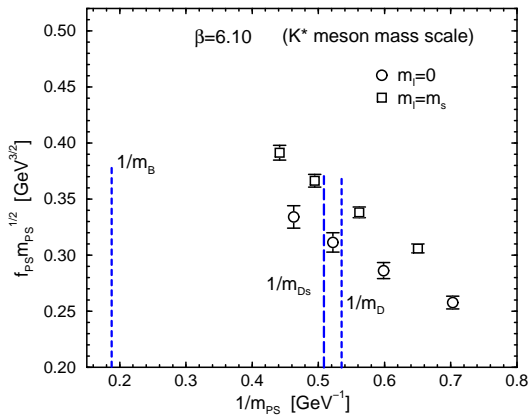,width=\figwidth}
\vspace{-1.0cm}
\caption{
The heavy-light decay constant multiplied by $\sqrt{m}$ in
physical units.}
\label{fig:decayconst}
\vspace{-0.2cm}
\end{figure}

For heavy quark, we use four values of $\kappa$
covering the physical charm quark mass with
$\gamma_F$ according to Eq.~(\ref{eq:fit_calib}).
As the light quark, we use three values of $\kappa$ with
$\gamma_F$ at the massless limit, which correspond to
the lightest three values used in \cite{Aniso01b}
for light hadron spectroscopy.
The meson correlator is calculated with the local meson
operator.
For the decay constant, we focus on the pseudoscalar mesons.
While one needs the matching constants to convert the lattice
result into the continuum theory, we only perform the
tadpole-improved tree-level matching in this paper.

We first extrapolate the result to the chiral limit, 
linearly in light pseudoscalar meson mass squared.
The physical ($u$, $d$) and $s$ quark masses are defined
with the massless limit and the $K$ meson mass.
Then they are fitted to the quadratic form in $1/m_{H}$,
$m_H$ the PS heavy-light meson mass, and interpolated to
the physical $m_D$ meson mass.
The decay constant of heavy-light meson is extrapolated
and interpolated in the form of $f_H\sqrt{m_H}$.
The result after chiral extrapolation is shown in
Figure~\ref{fig:decayconst}.
The gross feature of $m_H$ dependence is consistent with
previous works \cite{review}.
The masses and decay constants for physical quark masses
are listed in Table~\ref{tab:decayconst_latt}.

The hyperfine splittings are consistent with previous works,
while 35\% less than the experimental values.
On the other hand, the splitting $m_{D_s}-m_D$ is close to
the experimental value.
The $\beta$ dependence of decay constants are rather large, and
considered as of the discretization error and renormalization
effect, as well as the uncertainty in setting the scale on
quenched lattice.
Taking ratios of decay constants, $\beta$ dependence is largely
cancelled.
The result is consistent with the previous lattice works
\cite{review}.
For more quantitative discussion, we need to 
discuss more carefully on the systematic uncertainties due to
the anisotropy, and to include the matching of current with
continuum theory.
As our conclusion, the result of numerical simulation is encouraging
to pursuit more quantitative analysis
in this direction.


\begin{thebibliography}{99}

\bibitem{review}
 For recent review, 
  S.~Ryan, hep-lat/0111010.

\bibitem{NRQCD}
 B.~A.~Thacker and G.~P.~Lepage,
  Phys. Rev. D { 43} (1991) 196.

\bibitem{EKM97}
 A.~X.~El-Khadra, A.~S.~Kronfeld and P.B.~Mackenzie,
  Phys. Rev. D { 55} (1997) 3933.

\bibitem{NRimp} 
 M.~L\"uscher, S.~Sint, R.~Sommer, P.~Weisz and U.~Wolff,
  Nucl. Phys. B { 491} (1997) 323.

\bibitem{Aniso01a}
 J.~Harada,
  A.~S.~Kronfeld, H.~Matsufuru, N.~Nakajima and T.~Onogi,
  Phys. Rev. D { 64} (2001) 074501.

\bibitem{Aniso01b}
 H.~Matsufuru, T.~Onogi and T.~Umeda,
  Phys. Rev. D { 64} (2001) 114503.

\bibitem{Aniso02a}
 J.~Harada, H.~Matsufuru, T.~Onogi and A.~Sugita,
  in preparation.

\bibitem{Ume01}
 T.~Umeda, R.~Katayama, O.~Miyamura and H.~Matsufuru,
  Int. J. Mod. Phys. A { 16} (2001) 2215.

\bibitem{LM93}
 G.~P.~Lepage and P.~B.~Mackenzie,
  Phys. Rev. D { 48} (1993) 2250.

\bibitem{Kla98}
 T.~R.~Klassen,
  Nucl. Phys. B { 533} (1998) 557.

\end{thebibliography}
\end{document}